\documentclass[12pt]{article}
\pdfoutput=1
\usepackage{amsmath,amssymb,epsfig,ulem}
\usepackage{color}
\usepackage{multicol}

\allowdisplaybreaks

\newcommand{\nc}{\newcommand}
\nc{\beq}{\begin{equation}}
\nc{\eeq}{\end{equation}}
\nc{\barray}{\begin{eqnarray}}
\nc{\earray}{\end{eqnarray}}
\nc{\barrayn}{\begin{eqnarray*}}
\nc{\earrayn}{\end{eqnarray*}}
\nc{\bcenter}{\begin{center}}
\nc{\ecenter}{\end{center}}
\nc{\ket}[1]{| #1 \rangle}
\nc{\bra}[1]{\langle #1 |}
\nc{\0}{\ket{0}}
\nc{\mc}{\mathcal}
\nc{\er}[1]{(\ref{eq:#1})}
\nc{\onehalf}{\frac{1}{2}}
\nc{\partialbar}{\bar{\partial}}
\nc{\psit}{\widetilde{\psi}}
\nc{\Tr}{\mbox{Tr}}
\nc{\hc}{\mbox{H.c.}}
\nc{\ev}{\;\mathrm{eV}}
\nc{\mev}{\;\mathrm{MeV}}
\nc{\gev}{\;\mathrm{GeV}}
\nc{\tev}{\;\mathrm{TeV}}

\def\chii0{\chi_i^0}
\def\chij0{\chi_j^0}

\newcommand{\gsim}{\lower.7ex\hbox{$\;\stackrel{\textstyle>}{\sim}\;$}}
\newcommand{\lsim}{\lower.7ex\hbox{$\;\stackrel{\textstyle<}{\sim}\;$}}
\nc{\ttbar}{t\bar t}

\begin{document}

% 
% This file is to be included via
%                \input wgreport.tex 
% in Peskin's driverFile.tex.
% 
% 
% 
% 
\title{Kinematics of Top Quark Final States: A Snowmass White Paper}

%%%%%%%%%%%%%%%%%%%%%%%%%%%%%%%%%%%%%%%%%%%%%%%%%%%%%%%%%%%
%%%%%%%%%%%%%%%%%%%%%%%%%%%%%%%%%%%%%%%%%%%%%%%%%%%%%%%%%%%
%%%%%%%%%%%%%%%%%%%%%%%%%%%%%%%%%%%%%%%%%%%%%%%%%%%%%%%%%%%
%%%%%%%%%%%%%%%%%%%%%%%%%%%%%%%%%%%%%%%%%%%%%%%%%%%%%%%%%%%
\vspace{15pt}
\begin{center}
  \Large\bf 
Kinematics of Top Quark Final States:\\ A Snowmass White Paper
\end{center}
\vspace{5pt}
\begin{center}
{\sc Andreas Jung$^{a}$, Markus Schulze$^{b}$, and Jessie Shelton$^{c}$}\\
\vspace{10pt}
$^a$ {\it Fermilab, DAB5 - MS 357,\\ P.O. Box 500, Batavia, IL, 60510, USA}\\
\vspace{5pt} 
$^b$ {\it High Energy Physics Division,\\ Argonne National Laboratory, Argonne, IL 60439, USA }\\
\vspace{5pt} 
$^c$ {\it  Center for the Fundamental Laws of Nature, \\Harvard University, Cambridge, MA 02138, USA}\\

\vspace{10pt} {\it with contributions from}\\
 T.~Schwarz, S.~Berge, S.~Westhoff, and V.~Coco (for the LHCb Collaboration)

\end{center}

\begin{abstract} 
\vspace{2pt} 
\noindent 
This is the summary report of the Top Quark Kinematics working group
prepared for Snowmass 2013.  We study theoretical predictions for top pair differential
distributions, in both boosted and un-boosted regimes,
and present an overview of uncertainties and prospects for top spin correlations.
We study the prospects for measuring the inclusive SM top pair
production asymmetry $A_{FC}$ at LHC 14 as a function of systematic
error, and show that some improvement over current systematic
uncertainties, as customarily handled, is required for observing a
SM-size asymmetry.  Cuts on top pair invariant mass and rapidity do
not substantially alter this conclusion.  We summarize the conclusions
of contributed studies on alternate LHC measurements of the $t\bar t$
production asymmetry, in $t\bar t+$jet final states and in forward top
production at LHCb, both of which show good prospects for observing
SM-size asymmetries in 50 fb$^{-1}$ of data at LHC14.
\end{abstract}

%%%%%%%%%%%%%%%%%%%%%%%%%%%%%%%%%%%%%%%%%%%%%
\section{Introduction}
%%%%%%%%%%%%%%%%%%%%%%%%%%%%%%%%%%%%%%%%%%%%%

The top quark, the heaviest known elementary particle, has a lifetime significantly shorter than the
time scale required for hadronization.   Due to this short lifetime,
bare top quark properties can be observed by measuring the  kinematics of the top's decay products.
Top pair production events are important both as a signal in their own right and as a background to physics  beyond the Standard Model (BSM).  A detailed
 understanding of top quark kinematics is necessary in order to observe potential effects of BSM physics in top pair production, or to predict top backgrounds for  signals of direct BSM production. 

This white paper collects the results of the Top Kinematics working
group prepared for Snowmass 2013. It contains results of dedicated
studies performed by the coordinators of the working group (A.~Jung,
M.~Schulze, and J.~Shelton), with additional input from T.~Schwarz,
and summarizes the conclusions from contributed studies by S.~Westhoff and
S.~Berge and by the LHCb collaboration, which appear in full
elsewhere \cite{Coco:2013, Berge:2013csa}.

Section~\ref{sec:topkin-basicdistr} begins with a study of theoretical
uncertainties on basic kinematic distributions in top quark pair
production at the 14~and 33~TeV LHC, and in
Section~\ref{sec:topkin-boosted} the same uncertainties are considered
for the high $p_T$ tails of the top quark production cross-sections.
Section~\ref{sec:topkin-spincorrel} summarizes results on top quark
spin correlations. In Section~\ref{sec:topkin-afc}, the observability
of the inclusive SM top quark charge asymmetry at the 14 TeV LHC is
studied as a function of experimental systematic
uncertainties. 
Section~\ref{sec:topkin-newObs} summarizes studies for alternate
measurements of the top quark production asymmetry, both in $t\bar t+
$jet final states at LHC 14 and LHC 33 and inclusively at LHCb.
 
%%%%%%%%%%%%%%%%%%%
\section{Basic kinematic distributions}
\label{sec:topkin-basicdistr}
%%%%%%%%%%%%%%%%%%%

\begin{figure}[t]
  \centering
  \label{fig:topkin-basicdistr}
\includegraphics[width=0.32\hsize]{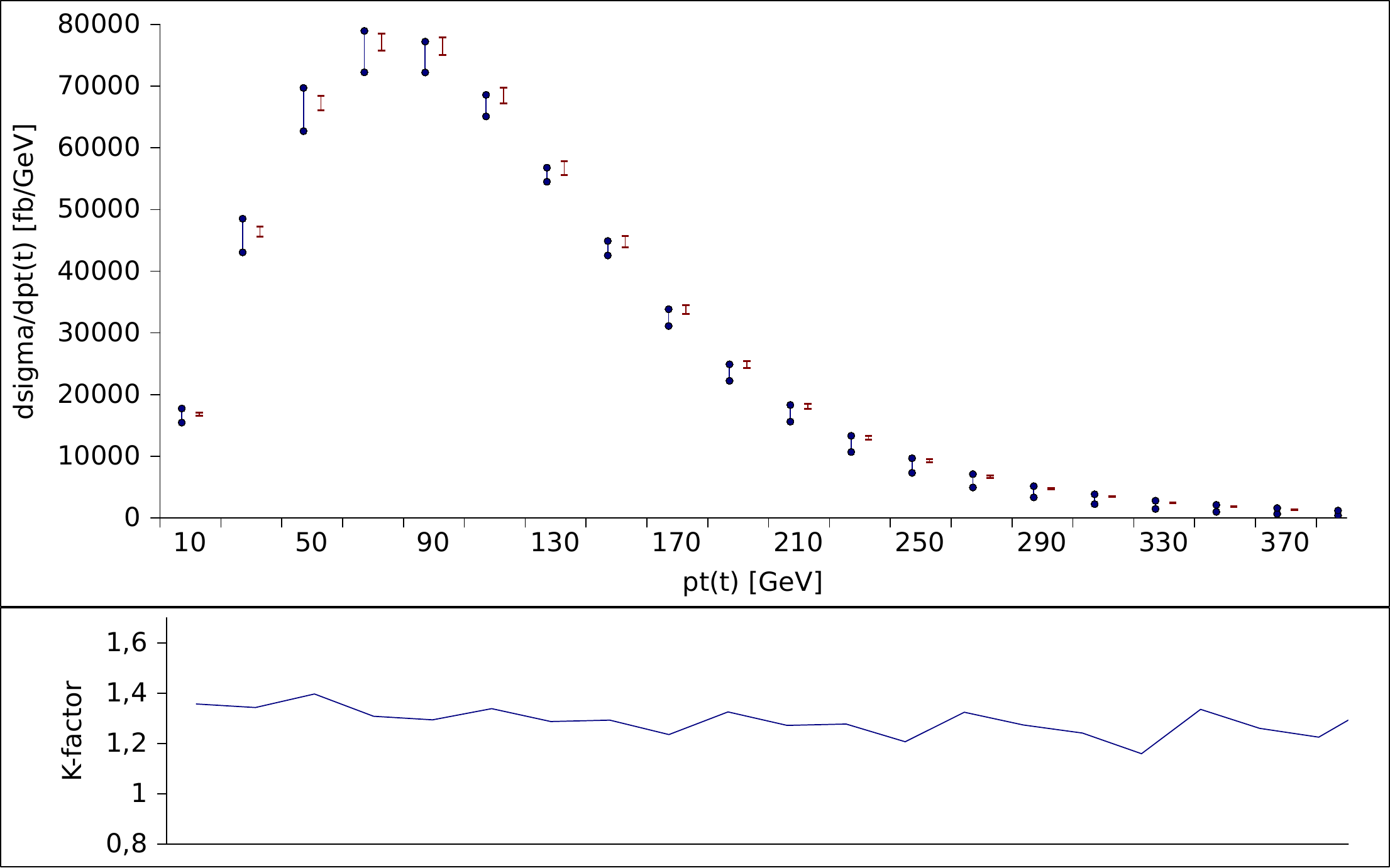}
\includegraphics[width=0.32\hsize]{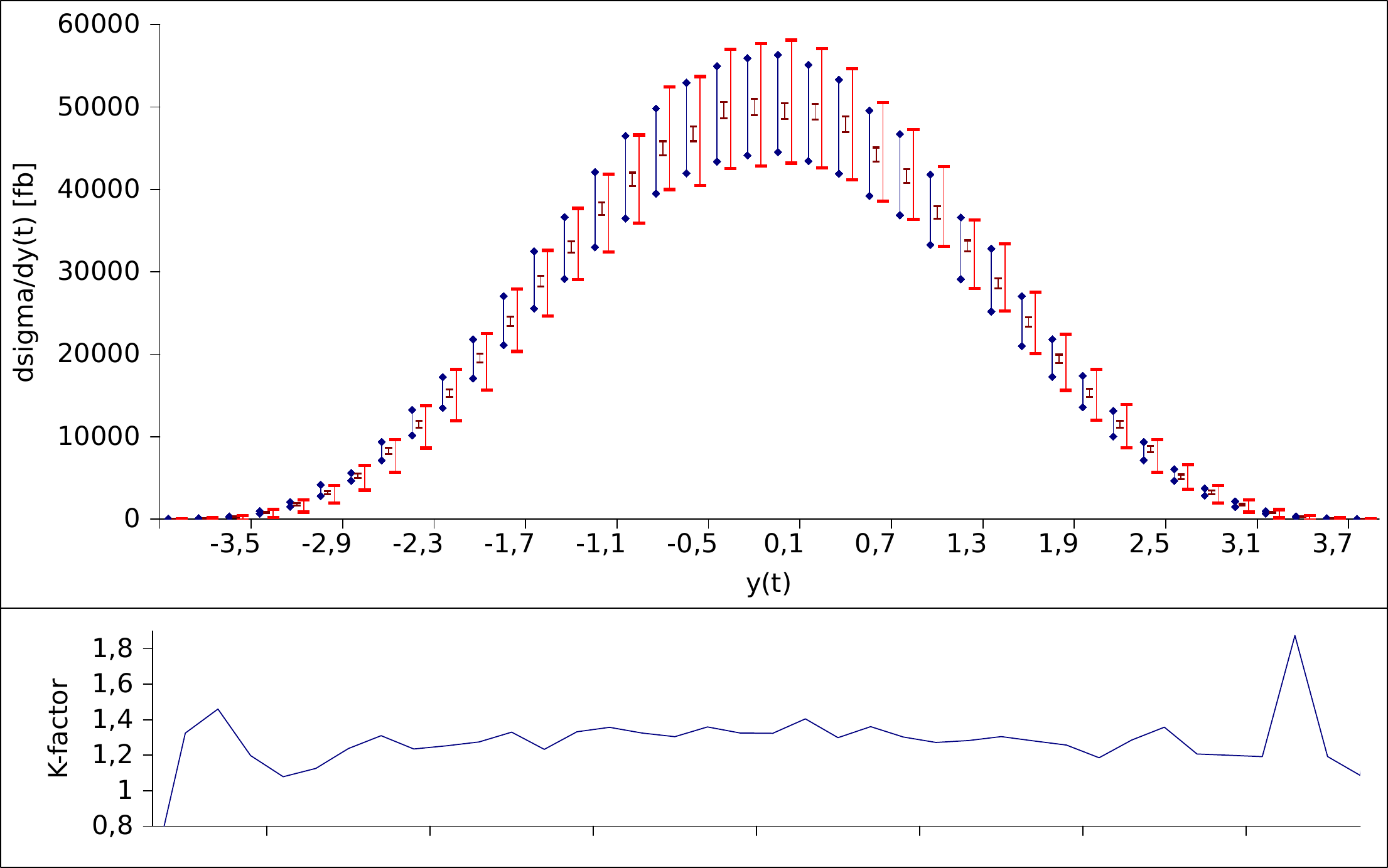}
\includegraphics[width=0.32\hsize]{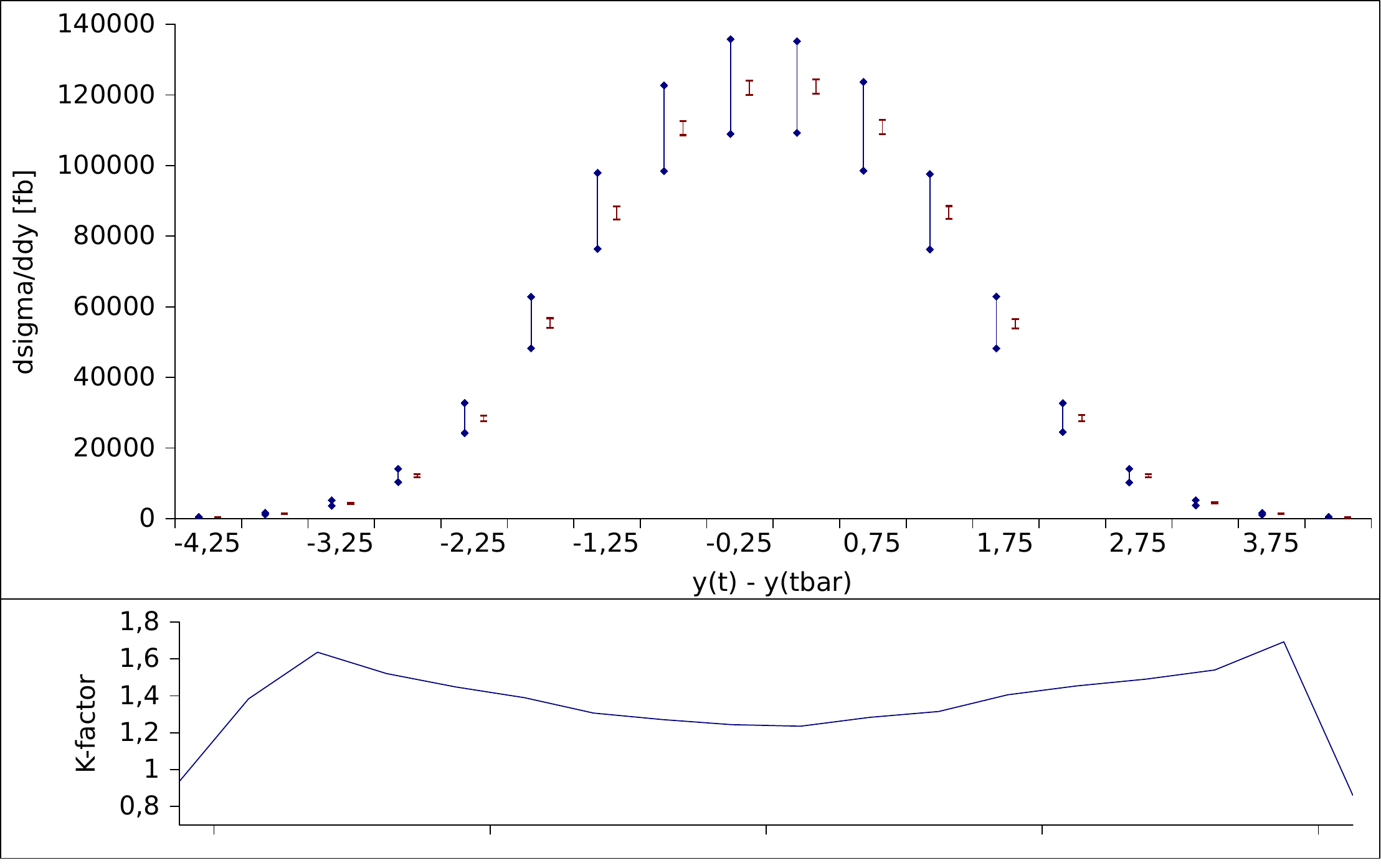}
\caption{NLO QCD predictions \cite{Campbell:2010ff} for transverse
  momentum, rapidity and rapidity difference at the $14$~TeV LHC.
  Blue error bars correspond to scale variation by a factor of two
  around $m_t$. Dark red error bands correspond to variation of
  different MSTW pdf error sets.  The additional band in the $y_t$
  distribution shows the variation with NN pdf error sets.}
\end{figure}

In this section we summarize theoretical uncertainty estimates for top
quark pair production at the 14~TeV and 33~TeV LHC.  The results are
obtained from the publicly available program
MCFM~\cite{Campbell:2010ff} at next-to-leading order (NLO) QCD.  Uncertainties are estimated
by varying renormalization and factorization scales as well as by
using different parton distribution functions and their error sets.
We assume stable top quarks and consider the total $t\bar t$ cross
section and distributions in the top quark transverse momentum, its
rapidity and the difference of top and anti-top quark rapidity.  The aim of this study
 is to provide 
% a unified overview of current predictions and
%  establish
error estimates that can be used to
extrapolate uncertainties on more complicated observables that involve
the top quark decay products.
Explicit NLO predictions for top-quark pair production and decay have been presented in
Refs.~\cite{Bernreuther:2001rq,Bernreuther:2004jv,Frixione:2006gn,Frixione:2007nw,Frixione:2007nu,Melnikov:2009dn,Bernreuther:2010ny,Denner:2010jp,Bevilacqua:2010qb,Campbell:2012uf,Denner:2012yc}.

The total NLO QCD $t\bar{t}$ cross section at the 14 TeV LHC is
$845$~pb.  Next-to-leading order QCD corrections reduce the scale
dependence by more than a factor of two and enhance the total cross
section by 30\%.  The residual scale uncertainty amounts to
about 12\% when varied by a factor of two around $m_t=173$~GeV.
Variations of 40 error sets of the MSTW pdf set~\cite{Martin:2009iq}
amount to shifts of about 2 \% of the total cross section at
leading and next-to-leading order QCD.  Using the different pdf fits
of the NN group~\cite{Ball:2012cx}, we find a total cross section
with a central value deviating by less than one \% from the MSTW
result.  However, the error bars of NN pdfs are significantly larger than 
those in the MSTW pdf sets, amounting to about 8 \%.

In Fig.~\ref{fig:topkin-basicdistr} we show the kinematic
distributions of top quark transverse momentum, rapidity and the
rapidity difference between top and anti-top quark at NLO QCD.  The
transverse momentum distribution can be predicted with an approximate
uncertainty of 15-20\%.  In contrast to the LO predictions, which have
an almost constant error over the entire $p_\perp$ range, NLO effects
show largest scale dependence close to threshold and smaller scale
dependence at higher energies.  Variations from different MSTW pdf
error sets amount to about a 2-8\% spread in the $p_\perp$ bins.  The
rapidity distribution of a top quark has a residual scale dependence
of up to 20\% in the central region.  In
Fig.~\ref{fig:topkin-basicdistr} we also show error bars from
variations of both MSTW and NN pdf error sets.  It is striking that
uncertainty estimates of the NN pdfs are significantly larger than the
ones from MSTW, reaching a similar size to the scale variation
band. Nevertheless, the central values agree well.  The difference between top
and anti-top rapidities is sensitive to the top quark charge asymmetry
at the LHC.  We find similar uncertainties as for the rapidity
distribution.  However, here, NLO effects introduce significant shape
changes with respect to the LO, as can be seen from the differential K-factor in the
lower pane of the plot.

Increasing the LHC center of mass energy from 14 TeV to 33 TeV
increases the total $t\bar t$ cross section by a factor of 6 to almost
5~nb. The residual scale uncertainty is 11\%, similar to the
14~TeV case.  Uncertainties from pdfs are expected to be larger since
the gluon large-$x$ pdfs is only weakly constrained.
The K-factor is notably smaller at $K=1.23$.

%%%%%%%%%%%%%%%%%%%%%%%%%%%%%%%
\section{Boosted kinematics}
\label{sec:topkin-boosted}
%%%%%%%%%%%%%%%%%%%%%%%%%%%%%%%

\begin{figure}[t]
  \centering
  \label{fig:topkin-boosted}
\includegraphics[width=0.45\hsize]{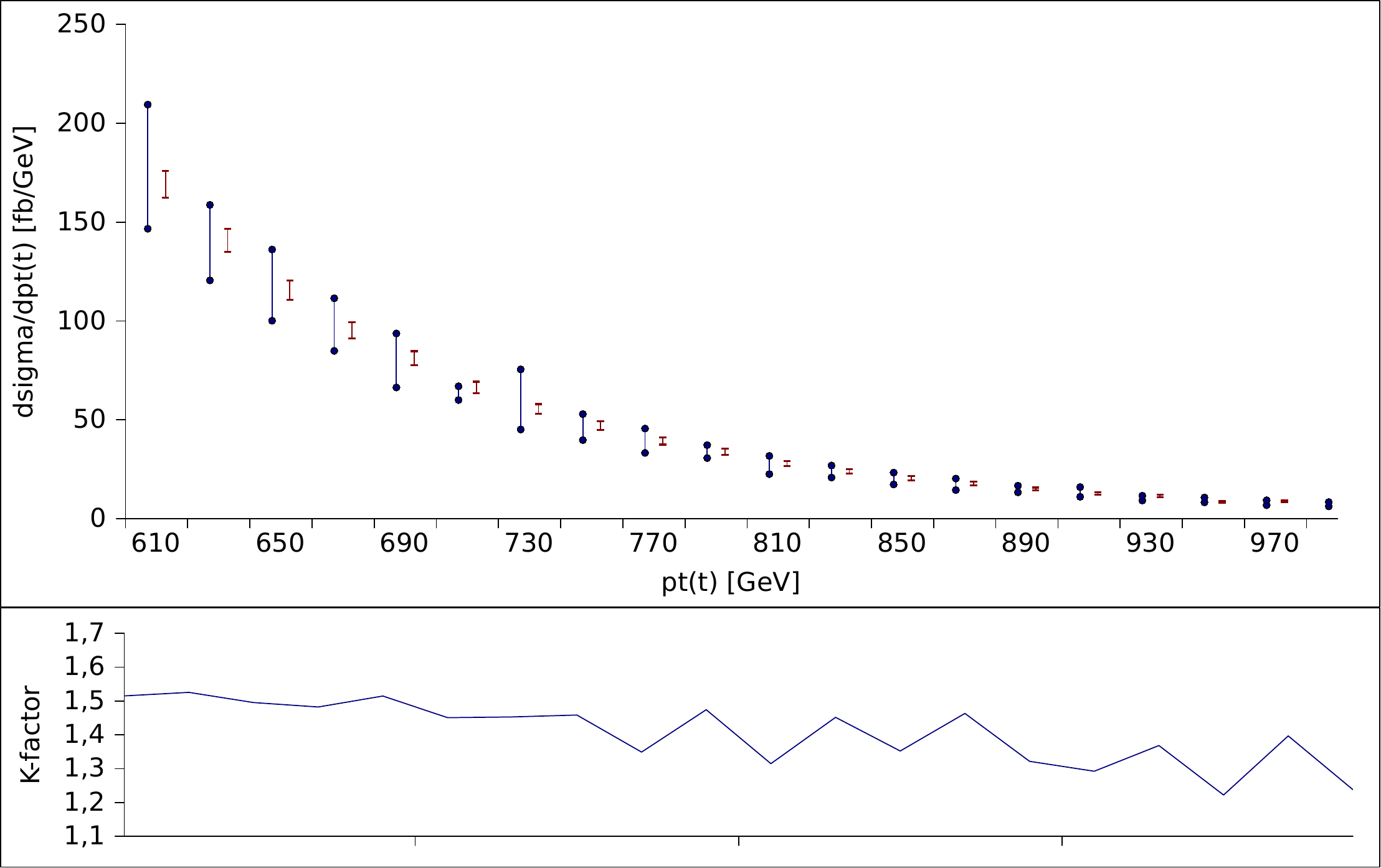}
\includegraphics[width=0.45\hsize]{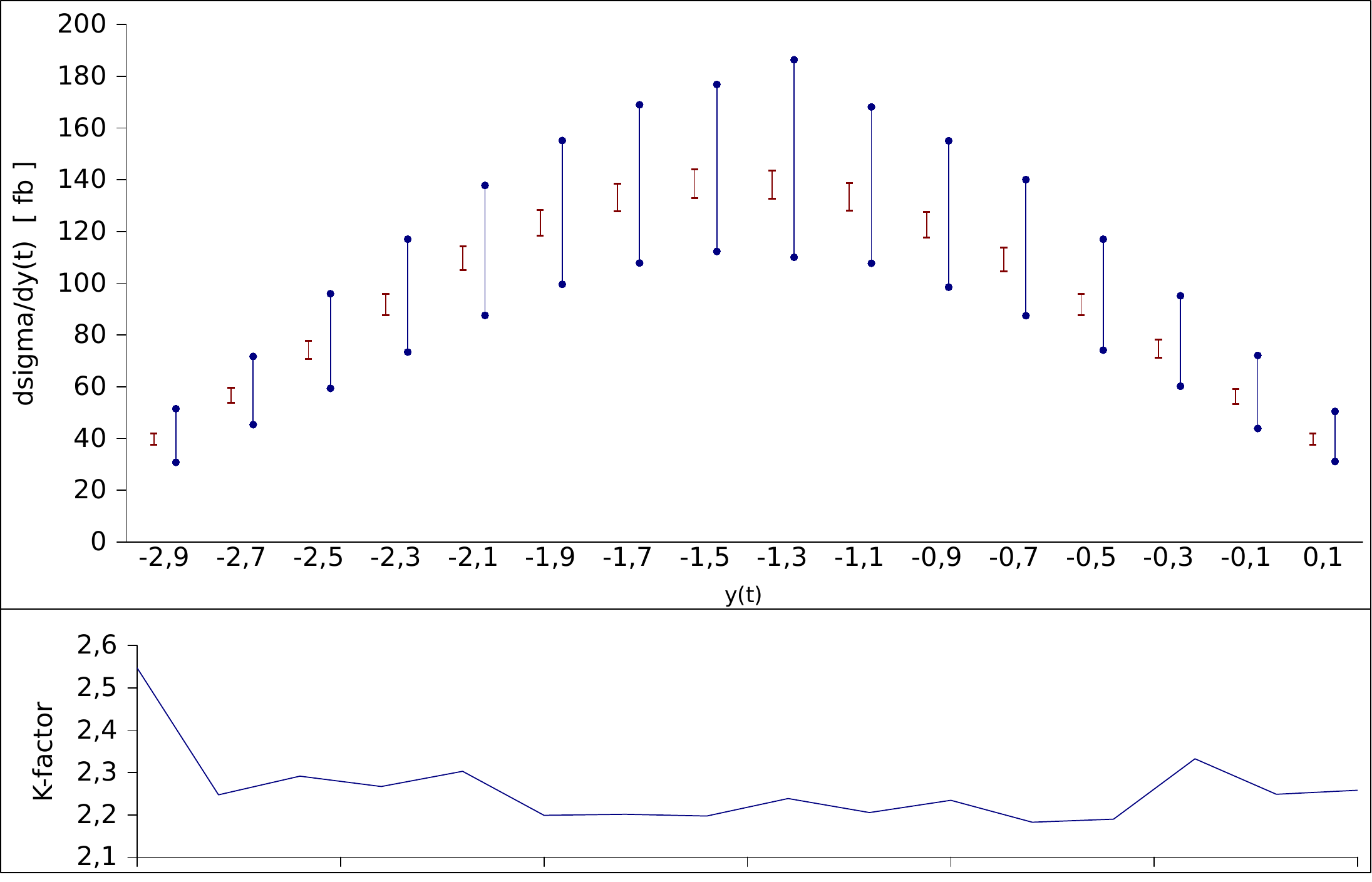}

\caption{NLO QCD predictions \cite{Campbell:2010ff} for top quark
  transverse momentum and rapidity with $p_\perp^{ t} \ge 600$~GeV at
  the $14$~TeV LHC.}
\end{figure}

In this section we summarize theoretical uncertainties of the
previously presented observables in the boosted regime at the 14~TeV
LHC.  In particular, we require that $p_\perp^t$ or $p_\perp^{\bar t}
\ge 600$~GeV.  The NLO $t\bar{t}$ cross section is $1.05$~pb
\cite{Campbell:2010ff}, about 800 times smaller than the total NLO
cross section.  Scale variations by a factor of two around a central
scale of $\mu_0=600$~GeV give an uncertainty of 15\%, similar to
the total NLO cross section.  Error bands of MSTW pdfs are about
10\%, twice as big as compared to the total cross section.  In
Fig.~\ref{fig:topkin-boosted}, we show transverse momentum and
rapidity distribution and their uncertainties for $p_\perp^t$ or
$p_\perp^{\bar t} \ge 600$~GeV.  The error bars of single histogram
bins range up to 20-30\%.  NLO QCD corrections induce moderate
shape changes but the overall K-factor of 1.45 is 10\% larger
than for the total cross section.

It is interesting to note that the calculations of
Refs.~\cite{Auerbach:2013by,Ferroglia:2013zwa} indicate that
soft-gluon effects can be important at high energies.  In particular,
Ref.~\cite{Auerbach:2013by} considers the $p_\perp$ distribution and
finds corrections of up to 100\% wrt. NLO results for highly
boosted top quarks.  Similarly, the authors of
Ref.~\cite{Ferroglia:2013zwa} calculate the soft plus virtual
approximation to NNLO QCD correction of the $t\bar{t}$ invariant mass
distribution and come to similar conclusions. For example, at
$m_{t\bar{t}} =1.5$~TeV their approximate NNLO corrections exceeds 
the size of the corresponding NLO QCD correction.  These results are in
apparent tension with the $\mathcal{O}(15-30\%)$ uncertainty estimates obtained in
NLO QCD calculations, and further work will be required to resolve this point.

The effects of weak virtual corrections in $t\bar{t}$ production have
been studied e.g. in Ref.~\cite{Kuhn:2013zoa} (see also references
therein).  Effects on the total cross section are small, at 2\%
for LHC energies of 14~TeV and 33~TeV.  The corrections can be
significantly larger in the tail of energy-related differential
distributions.  As shown in Ref.~\cite{Kuhn:2013zoa}, at transverse
momenta $p_\perp^t \approx 1$~TeV weak virtual corrections are
$-10$\%  with respect to the LO prediction, and they grow to $-18$\% at
$p_\perp^t \approx 2$~TeV.  Similarly at invariant masses of
$m_{t\bar{t}}\approx 2$~TeV the corrections amount to $-6$\%.
Effects of partial cancellations between these weak virtual
corrections with the corresponding emission of collinear $Z$ and $W$
bosons has been studied in Ref.~\cite{Baur:2006sn}.  It was found that
at $p_\perp^t = 500$~GeV the cancellation is 1-2\%.  At higher
energies $p_\perp^t \approx 1$~TeV the cancellation can become larger,
but the details of acceptance cuts may significantly affect the
relative importance of the weak boson emission processes.

%%%%%%%%%%%%%%%%%%%%%%%%%%%%%
\section{Top quark spin correlations}
\label{sec:topkin-spincorrel}
%%%%%%%%%%%%%%%%%%%%%%%%%%%%%

\begin{figure}[t]
  \centering  
\label{fig:topkin-spincorr}
\includegraphics[width=0.25\hsize]{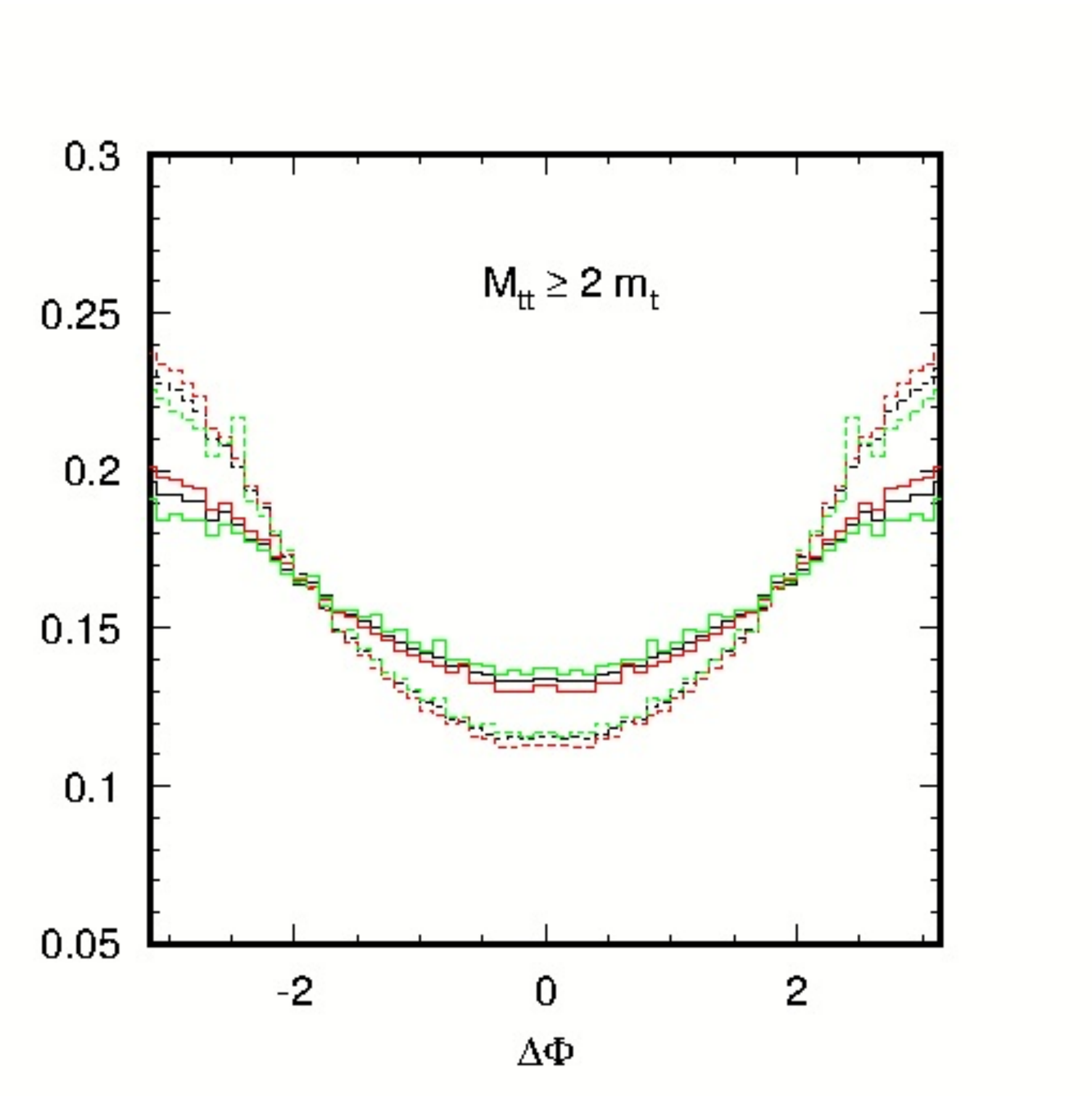}
\includegraphics[width=0.32\hsize]{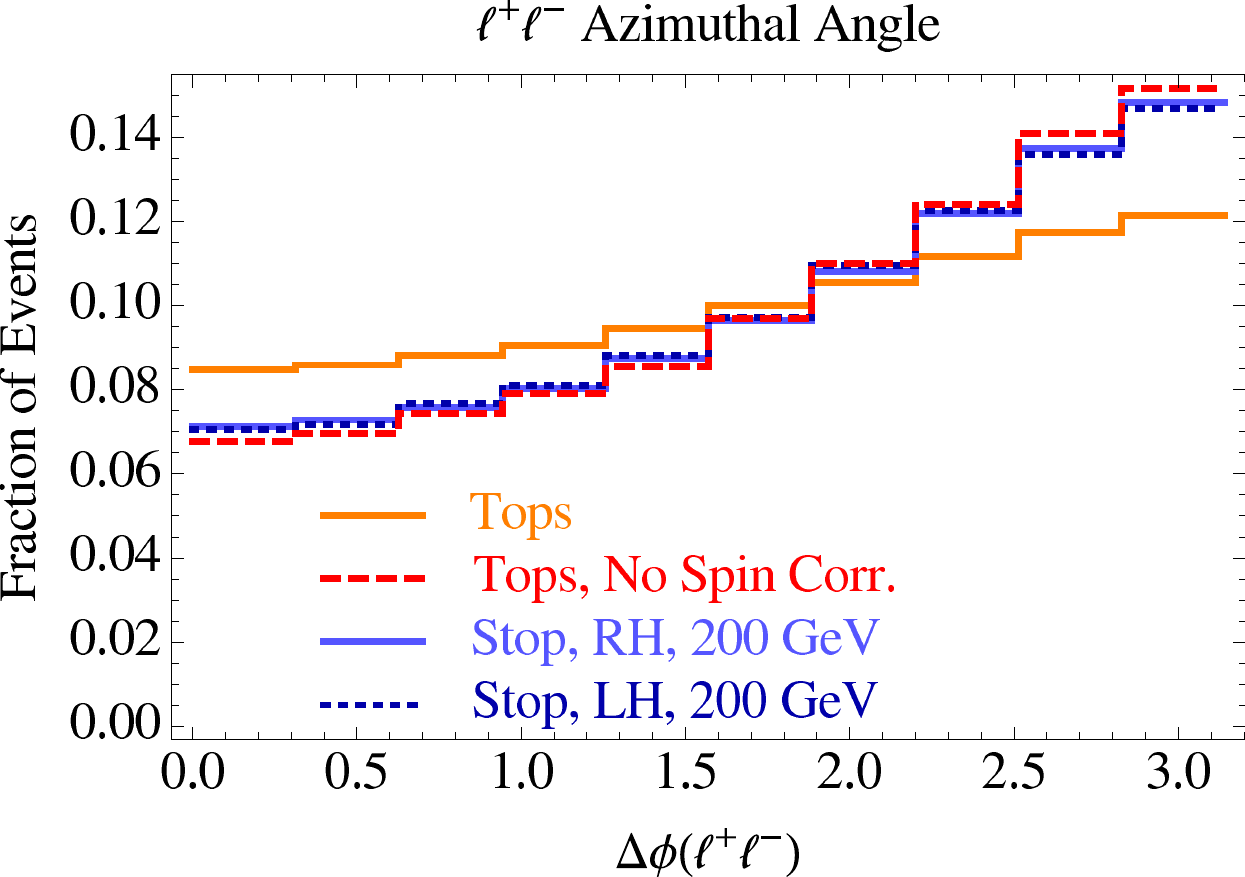}
\includegraphics[width=0.32\hsize]{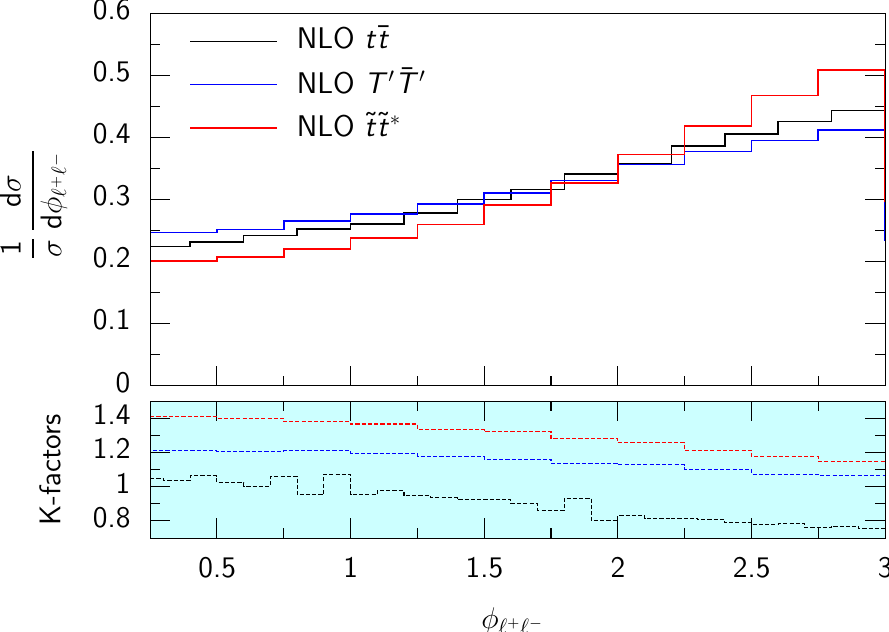}
\caption{ Distributions of the dilepton azimuth angle from
  Refs.~\cite{Bernreuther:2013aga,Han:2012fw,Boughezal:2013xx},
  respectively.  Fig.~(a) shows spin correlated (solid) and spin
  uncorrelated (dashed) top quarks. The hatched area around the solid
  lines corresponds to scale variations.  Fig.~(b) compares SM $t\bar
  t$ spin correlations with 200~GeV stop quarks decaying into a
  massless neutralino and a top quark.  In (c) NLO QCD predictions are
  shown for SM $t\bar t$ (black), fermionic partners of mass 500~GeV
  (blue) and scalar partners of mass 250~GeV (red).  The corresponding
  K-factors are shown in the lower pane.  }
\end{figure}

Top quark spin correlations are a unique tool for studying the
interplay between electroweak and strong physics in the top quark
sector. After the evidence of top quark spin correlations at the
Tevatron \cite{Abazov:2011gi} and recently the observation at the LHC
\cite{ATLAS:2012ao,CMS:ttbspincorrel}, experimental analyses will soon
be able to probe the effects of New Physics models on SM spin
correlations.  

The cleanest $t\bar t$ samples are the ones with two opposite sign
leptons in the final state.  Spin correlations in this di-leptonic
decay mode manifest themselves most prominently in the angle between
the two leptons.  In fact, the azimuthal opening angle has been shown
to be most robust under higher order corrections and parton showering
effects.  The effects of higher order corrections have been studied
e.g. in Refs.~\cite{Melnikov:2009dn,Bernreuther:2010ny}.  For standard
acceptance cuts, NLO QCD effects introduce shape changes of at most
20\%.  If additional cuts are applied that enhance spin
correlations, NLO corrections increase the correlation even further.
Electroweak corrections have negligible effects, and scale variations
are vanishingly small because distributions are typically normalized
(see Fig.~\ref{fig:topkin-spincorr}~a).  On the experimental side, the
reconstruction of the lepton opening angle in the laboratory frame
does not involve the kinematics of other particles and can therefore be
extracted with small systematic uncertainties.  The normalized
azimuthal opening angle distribution is therefore an ideal observable
for studying top quark spin correlations.  Of course, other
constructions such as helicity angles, double differential
distributions, and asymmetries can also be explored.

It has been shown that top quark spin correlations can be used to
distinguish SM top quarks from scalar or fermionic partners.  For
example, using spin correlations alone light stop quarks of
$m_{\tilde{t}} = 200$~GeV can be excluded at 95\% C.L.  using
20~$\mathrm{fb}^{-1}$ at the 8 TeV LHC \cite{Han:2012fw}.  It is also
possible to use spin correlations for distinguishing heavy fermionic
top partners from scalar partners.  It has been pointed out
\cite{Boughezal:2012zb,Boughezal:2013xx} that QCD corrections play an
important role in correctly understanding these processes.  In fact,
for the pair production of light scalar top partners with $m_{\tilde
  t} \le 300$~GeV and heavy fermion partners with $m_{T^\prime} \ge
500$~GeV (decaying into $t\bar t$ final states plus large missing
energy), significantly different NLO $K$-factors can result in
approximately equal cross sections $\sigma^\mathrm{NLO}_{\tilde t
  \tilde t^*} \approx \sigma^\mathrm{NLO}_{T \bar T^\prime}$.  Spin
correlations, however, show significantly different shapes and help
to separate the two hypotheses. Fig.~\ref{fig:topkin-spincorr}~c shows
an example for the processes $pp\to\tilde t \tilde t^* (m_{\tilde
  t}=250~\mathrm{GeV}) \to t \bar t + \chi^0 \chi^0
(m_{\chi^0}=50~\mathrm{GeV})$ and $pp\to T \bar T^\prime (m_{\tilde
  t}=500~\mathrm{GeV}) \to t \bar t + A^0 A^0
(m_{\chi^0}=50~\mathrm{GeV})$.  It should be noted that NLO
$K$-factors (lower pane of Fig.~\ref{fig:topkin-spincorr}~c) are
similar in size to the separation between SM $t\bar t$ production
and the BSM signals, emphasizing the importance of higher order
effects for these processes.  

In the event of a discovery of a new
resonance which decays into $t\bar t$ pairs, top quark spin
correlations can also be used to analyze the couplings of this new
particle \cite{Baumgart:2011wk,Caola:2012rs}.  Another interesting
aspect are New Physics contributions to the top quark chromomagnetic
$\hat\mu_t$ and electric $\hat d_t$ dipole moments.
Refs.~\cite{Baumgart:2012ay,Bernreuther:2013aga} demonstrate that New
Physics contributions to $\hat\mu_t$ and $\hat d_t$ can be exposed
through spin correlations in the di-leptonic and in the semi-leptonic
decay mode.  From the dileptonic sample of the $20~\mathrm{fb}^{-1}$
run at 8~TeV, it should be possible to constrain
$\mathrm{Re}\,\hat\mu_t$ and $\mathrm{Re}\,\hat d_t$ at the few percent
level.  The imaginary parts $\mathrm{Im}\,\hat\mu_t$ and
$\mathrm{Im}\,\hat d_t$ can be constrained from lepton-top helicity
angles in the semi-leptonic channel where a full reconstruction of the
$t\bar t$ system is possible. Using the same dataset limits of about
15-20\% are possible.  Ref.~\cite{Baumgart:2012ay} finds that
constraints of 1\% or below are possible with
$100~\mathrm{fb}^{-1}$ at 13~TeV.

%%%%%%%%%%%%%%%%%%%%
\section{Inclusive Top Quark Charge Asymmetry at LHC 14}
\label{sec:topkin-afc}
%%%%%%%%%%%%%%%%%%%%

Here we detail some estimates toward the measurability of the SM
forward-backward asymmetry $A_{FC}$ at the 14 TeV LHC.  The increased
center of mass energy, relative to LHC 7 and LHC 8, increases the
proportion of $t\bar t$ events that arise from (symmetric) gluon
fusion, so the size of the signal decreases with increasing center of
mass energy.  As we will demonstrate below, the observability of the
SM asymmetry at LHC 14 depends sensitively on the evolution of
experimental systematic uncertainties, and we will discuss prospects
and scenarios for their improvement.

Already at LHC 7, measurements of the top forward-central asymmetry
are systematically limited.  Current LHC measurements in the
lepton$+$jets channel are done with the full 5 fb$^{-1}$ 7 TeV
dataset. CMS finds \cite{Chatrchyan:2012cxa}
\begin{equation}
\label{eq:systref}
A_{FC}=0.004\pm 0.010\pm 0.011 .
\end{equation}
ATLAS, in their similar study, marginalizes over sources of
systematic uncertainty according to a novel procedure
\cite{Choudalakis:2012hz} and claims \cite{ATLAS-CONF-2013-078}
\begin{equation}
A_{FC}= 0.006\pm 0.010
\end{equation}
where the quoted uncertainty is now almost entirely statistical in
origin.  Prior to the marginalization, the individual systematic
uncertainties tabulated in Table 3 of Ref.~\cite{ATLAS-CONF-2013-078}
add in quadrature to a total systematic uncertainty comparable to that
of CMS' result.  Without this marginalization procedure, the
systematic uncertainty on the current measurements of $A_{FC}$ is
comparable to the size of the predicted SM effect
\cite{Bernreuther:2012sx},
\begin{equation}
A_{FC}= 0.0123 \pm 0.0005
\end{equation}
at LHC 7.

\begin{figure}
  \centering
  \label{fig:topkin-SMAsymmPred}
\includegraphics[width=0.45\hsize]{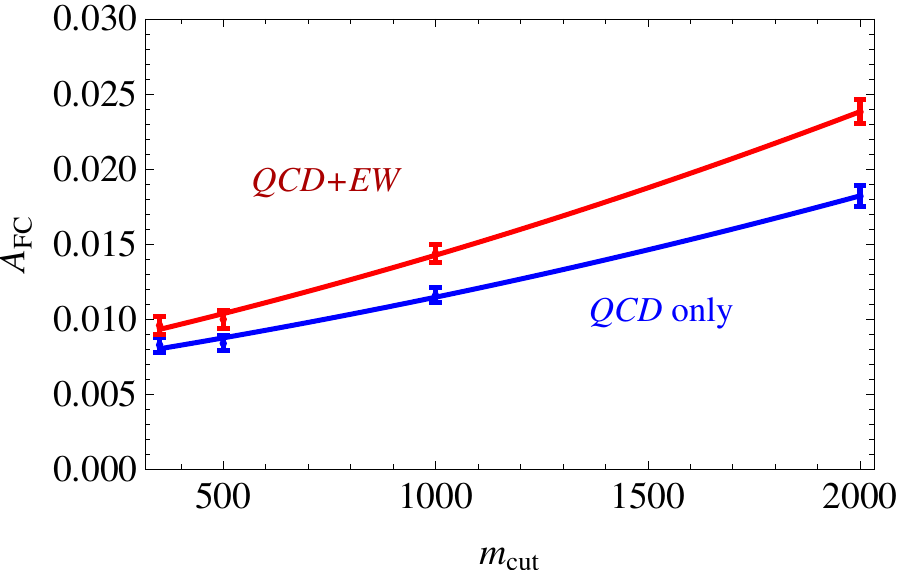}
\includegraphics[width=0.45\hsize]{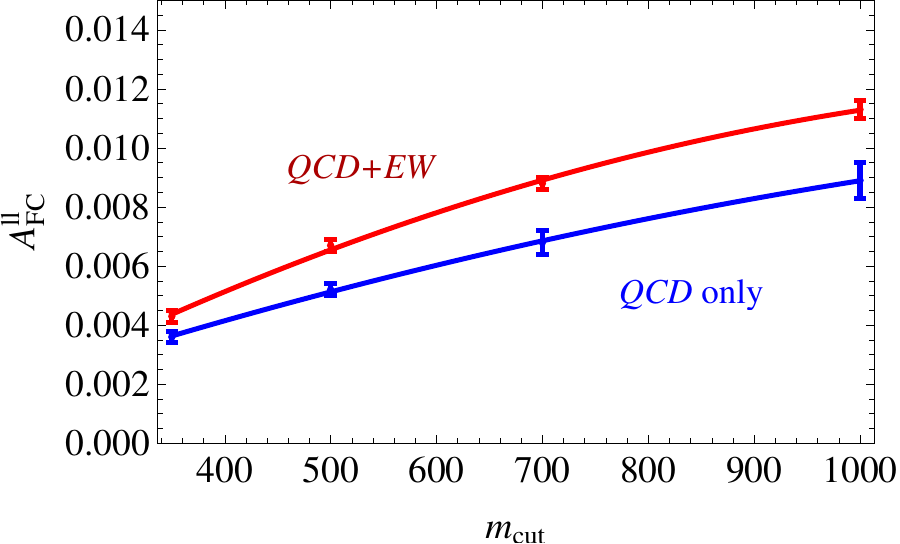}
\caption{Inclusive NLO SM predictions for the top quark asymmetry
  (left) and {\it inclusive} lepton asymmetry (right) as a function of
  minimum CM invariant mass.  Points with error bars are results
  calculated in Ref.~\cite{Bernreuther:2012sx}, showing the theoretical
  uncertainty arising from scale variation; the curves are
  interpolations used in the present study. Incorporating standard
  acceptance cuts on the leptons reduces the lepton asymmetry by more
  than a factor of two.}
\end{figure}

SM predictions for LHC 14 as a function of cuts on minimum top pair
invariant mass $m_{t\bar t}$ or velocity $\beta_{t\bar t}$ are
calculated in Ref.~\cite{Bernreuther:2012sx} and shown in
Fig.~\ref{fig:topkin-SMAsymmPred}.  Specifically, we use predictions
for the quantity
\begin{equation}
A_{FC}^{\eta} = \frac{N(\Delta|\eta| > 0)-N(\Delta|\eta| < 0)}{N(\Delta|\eta| > 0)+N(\Delta|\eta| < 0)}
\end{equation}
where $\Delta|\eta|\equiv |\eta_t |-|\eta_{\bar t}| $ looks at whether
the reconstructed top or anti-top is more central according to
lab-frame {\it pseudo-rapidity}.  Cutting on either CM invariant mass
or CM rapidity increases the proportion of $q$-$\bar q$-initiated top
pair events relative to gluon-initiated events, and thus enhances the
signal.  It can be seen from Fig.~\ref{fig:topkin-SMAsymmPred},
however, that even after imposing kinematic cuts, the size of the
signal at the 14 TeV LHC is comparable to the systematic uncertainties
on the current measurements.  Unlike at the 7 TeV LHC, statistical
uncertainties will not be limiting given current projections for the
integrated luminosities to be achieved at 14 TeV.
%%%%
%%%%

The dominant contributions to the experimental systematics are collected
in Table~\ref{tab:afcsys}.  Several systematic uncertainties arise
from uncertainties associated with the nature of the measurement of
the forward charge asymmetry, as well as assumptions in background and
top quark modeling.  Each uncertainty is treated in its own way, but
the common technique for $A_{FC}$ is to compare a measurement using a
baseline model with one where one of the inputs has been varied within
its known error.

Systematic uncertainties taken into account in measurements of
$A_{FC}$ include those from detector effects (such as jet energy
scale/resolution, lepton energy scale/resolution, and pileup),
modeling uncertainties (such as MC generator and hadronization,
MC-derived backgrounds, and PDF uncertainties), and uncertainties from
measurement techniques (such as unfolding).  The leading two systematics in each
measurement are either simulation modeling uncertainties or
calibration uncertainties for leptons and jets.  These uncertainties
will not necessarily scale with integrated luminosity without
improvements in technique.  Jet energy and lepton energy scale, which
are derived from comparisons between data and Monte Carlo, could
possibly be reduced in a larger dataset, though uncertainties are
currently dominated by disagreement between Monte Carlo generators.
Moreover, any increased pileup will likely contribute to increasing the
overall uncertainty.  However, jet and lepton energy scale
uncertainties are already small enough to allow a significant
measurement of $A_{FC} > 0.01$ in the semi-leptonic channel if
$t\bar{t}$ modeling uncertainties can be significantly reduced, at
least by half.  One approach to reducing the limiting modeling uncertainties has already been advanced by ATLAS in their most
recent semi-leptonic measurement~\cite{ATLAS-CONF-2013-078}.

Additionally, combining measurements across experiment and channel
will further reduce uncorrelated uncertainties, allowing for
reductions in the overall systematic uncertainty. Given that systematic uncertainties
on $A_{FC}$ are already on the order of 0.01, it is certainly
reasonable to assume that a significant measurement of the SM forward
charge asymmetry will be possible at LHC 14, and will grow in
significance for physics beyond the Standard Model which predicts an
enhancement.  However, it will largely be improvements in modeling
techniques that will allow this, not increased integrated luminosity,
and no further benefit is obtained at the HL-LHC.
%%%%%%%%%%%%%%%%%%%%%%%%%%%%%%%%%%%
\noindent
\begin{table}[t]
\begin{footnotesize}
\begin{center}
\begin{tabular}{| c l | c c |}
\hline
\hline
  Experiment & Channel  & Leading Systematic & Second Leading Systematic \\
  \hline
  ATLAS & Semi-leptonic& Jet energy scale & Lepton energy scale\\
  &  & 0.003 & 0.003\\
 & Dileptonic  & $t\bar{t}$ model dependence & Multi-jet modeling\\
    & & 0.015 & 0.012\\
  \hline
    CMS & Semi-leptonic & $t\bar{t}$ model dependence & Lepton energy scale\\
      &  & 0.007 & 0.006\\
 & Dileptonic  & Migration matrix (+Sys.) & $t\bar{t}$ model dependence (-Sys)\\
   &  & +0.01 & -0.03\\
 \hline
\end{tabular}
\end{center}
\end{footnotesize}
\caption{Measurements of $A_{FC}$ using approximately 5 $\mathrm{fb^{-1}}$ of 7 TeV data 
  at CMS and ATLAS for both $t\bar{t}$ decay channels.  The two largest systematics are 
  shown along with their respective magnitudes.  Note that the dilepton measurement at ATLAS 
  evaluates systematics separately for each lepton channel and combines them to reduce the 
  overall systematic uncertainty; the listed systematic uncertainties are for the 
  electron-electron channel.  The migration matrix uncertainty, as listed in the CMS dilepton 
  measurement, is due to the finite Monte Carlo statistics used in the response 
  matrix of an unfolding technique.\label{tab:afcsys}}
\end{table}
%%%%%%%%%%%%%%%%%%%%%%%%%%%%%%%%%%%%
%%%%
\begin{figure}
  \centering
  \label{fig:topkin-SMAsymmProj}
\includegraphics[width=0.45\hsize]{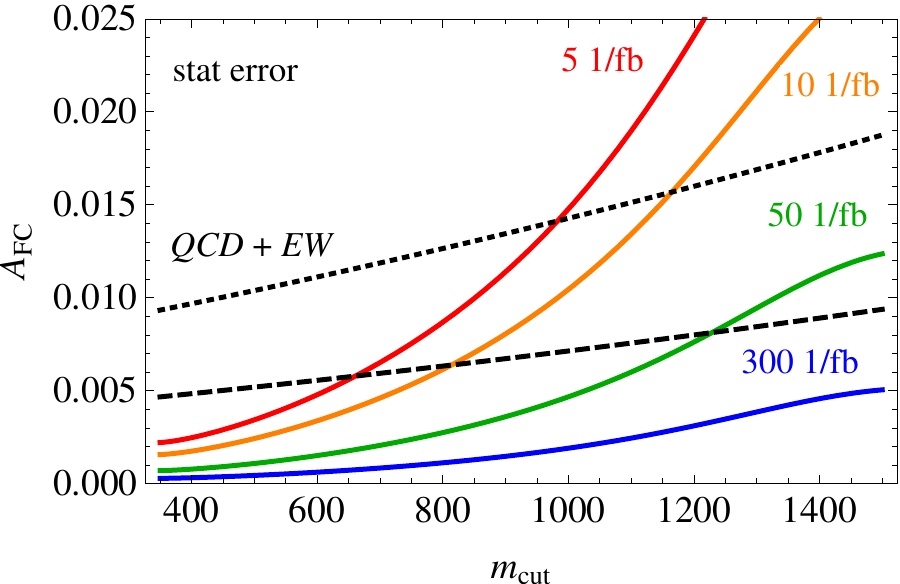}
\includegraphics[width=0.45\hsize]{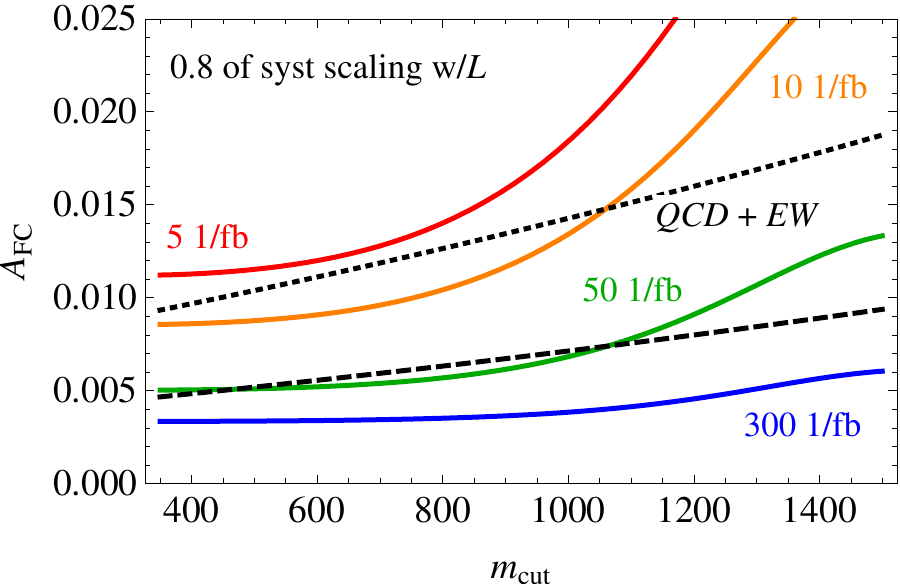}\\
\includegraphics[width=0.45\hsize]{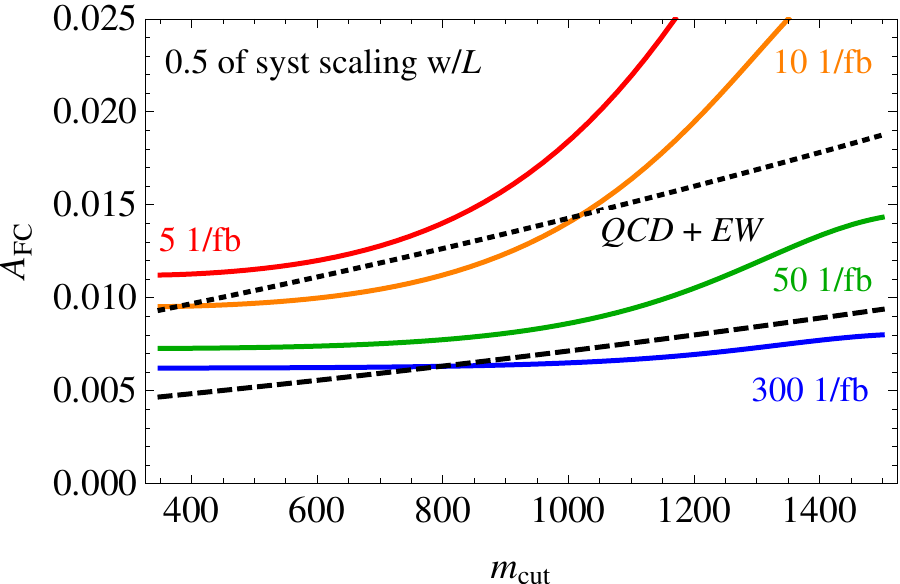}
\includegraphics[width=0.45\hsize]{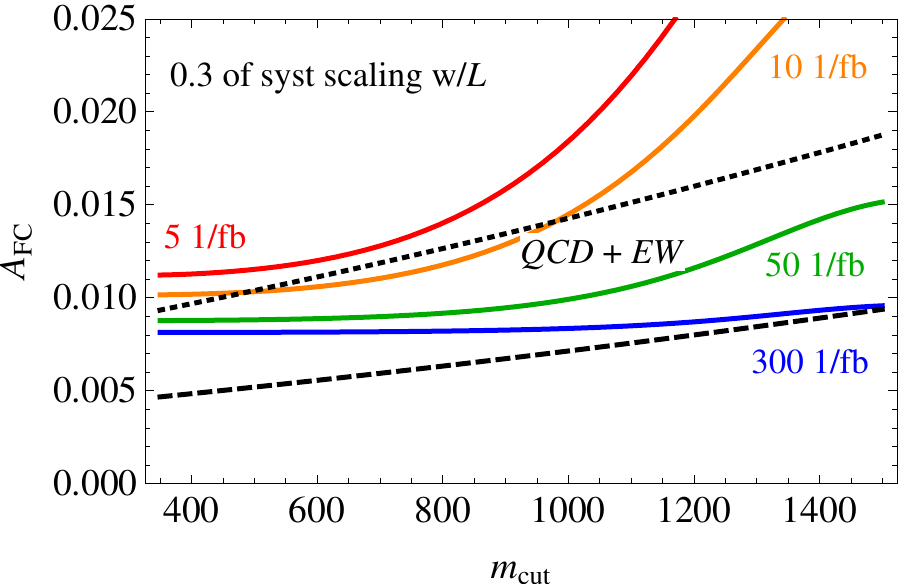}
\caption{ SM asymmetry compared to projected uncertainty on the
  measurement in the lepton$+$jet channel at LHC 14, as a function of
  minimum $m_{t\bar t}$, with three different scenarios for the
  improvement of systematic uncertainties with luminosity.  The dotted
  line shows the SM (QCD and EW combined) predictions of
  Ref.~\cite{Bernreuther:2012sx}, and the dashed line shows 0.5
  $\times$ the asymmetry to indicate $95\%$ CL sensitivity. Top left:
  statistical error only.  Top right: statistical error combined in
  quadrature with systematic error given by Eq.~\ref{eq:systref} at 5
  fb$^{-1}$, with 0.8 of the systematic uncertainty scaling as
  $1/\sqrt{\mathcal{L}}$.  Bottom left: statistical error combined in
  quadrature with systematic error given by Eq.~\ref{eq:systref} at 5
  fb$^{-1}$, $0.5$ of which scales as $1/\sqrt{\mathcal{L}}$.  Bottom
  right: statistical error combined in quadrature with systematic
  error given by Eq.~\ref{eq:systref} at 5 fb$^{-1}$, $0.3$ of which
  scales as $1/\sqrt{\mathcal{L}}$. }
\end{figure}
%%%%

We present three scenarios for potential improvement of
systematic uncertainties to show the sensitivity range
at LHC 14 that results.  
In Fig.~\ref{fig:topkin-SMAsymmProj} we plot statistical and
combined statistical$ +$systematic uncertainties on $A_{FC}$ at LHC 14.  Statistical errors are shown assuming semi-leptonic top
pair decays with an approximate (flat) efficiency for top
identification and reconstruction of $\epsilon = 0.2$ per
(semi-leptonic) event.  This efficiency represents the requirement
that at least one (isolated) lepton and at least 4 jets lie within
$|\eta|< 2.4$, with $p_{T,j}> 30$ GeV, $p_{T,e}> 30$ GeV, and
$p_{T,\mu}> 20$ GeV, with the further requirement that solving
quadratic equations for the missing neutrino four-momentum yield a
physical solution.  The cross-section is normalized to a NLO
prediction of 845 pb using a constant K-factor.  In all
cases we take the systematic uncertainty at 5 fb$ ^ {-1} $ to be given
by the systematic uncertainty reported by CMS with 5 fb$ ^ {-1} $ of
LHC 7 data, and then evolve the systematic uncertainties with
luminosity assuming a fraction $(0.3, 0.5, 0.8) $ of the systematic
uncertainty scales as $1/\sqrt{\mathcal{L}}$.  Based on this estimate,
with sufficient luminosity, LHC14 will have sensitivity to the SM
asymmetry if at least $0.5$ of the systematic errors scale with
luminosity.   Optimistic scenarios involving significant improvement in modeling
techniques would more closely resemble the ``statistical only'' cases, and would have excellent prospects
with much less luminosity.

%%%%%%%%%%
\subsection{Leptonic charge  asymmetry}
%%%%%%%%%%

The dileptonic channel offers another window into top pair production
asymmetries, albeit at smaller statistics due to the reduced branching
fraction.  Current LHC measurements in the dileptonic channel are done
 at 7 TeV and have, with 5 fb$^{-1} $, comparable systematic errors to
those in the semileptonic channel \cite{ATLAS:2012sla, CMS:iya}.
An alternate possibility in the dileptonic channel is to measure an
asymmetry of the leptons rather than of the tops themselves,
\begin{equation}
\label{eq:lepasymm}
A_{\ell\ell} = \frac{N(\Delta|\eta_\ell| > 0)-N(\Delta|\eta_\ell| < 0)}
    {N(\Delta|\eta_\ell| > 0)+N(\Delta|\eta_\ell| < 0)},
\end{equation}
where $\Delta |\eta_\ell|\equiv |\eta_{\ell^+}|-|\eta_{\ell ^-} |$.
The systematic uncertainty quoted in the measurement of this leptonic
asymmetry is smaller than the uncertainty on the parent top asymmetry,
$0.6\%$ in CMS \cite{CMS:iya} and $0.8\%$ in ATLAS
\cite{ATLAS:2012sla}.  However, the asymmetry itself is also smaller,
especially when realistic geometric acceptance cuts are taken into
account.  We emphasize that including standard acceptance cuts
decreases the size of the lepton asymmetry by more than a factor of
two \cite{Bernreuther:2012sx}.

%%%%
\begin{figure}
  \centering
  \label{fig:topkin-SMAsymmProjDilep}
\includegraphics[width=0.45\hsize]{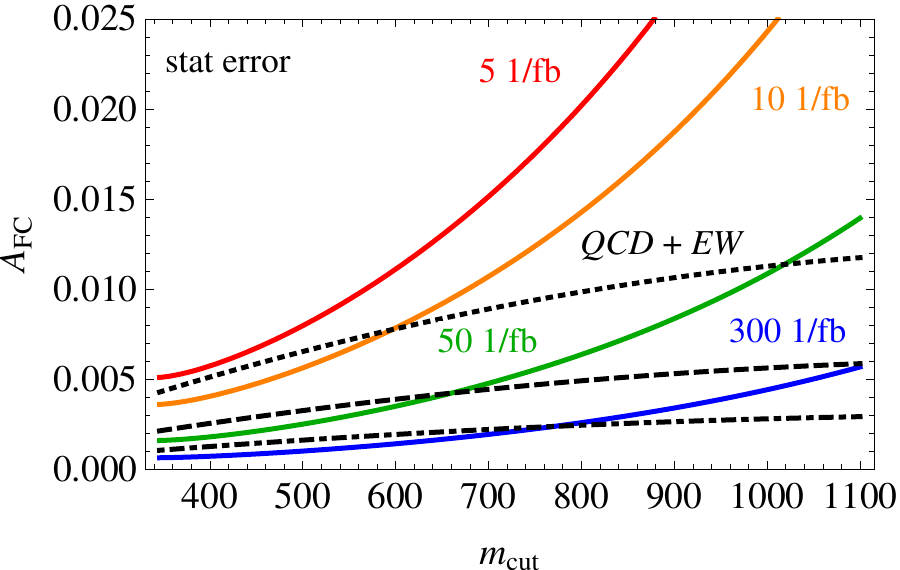}
\includegraphics[width=0.45\hsize]{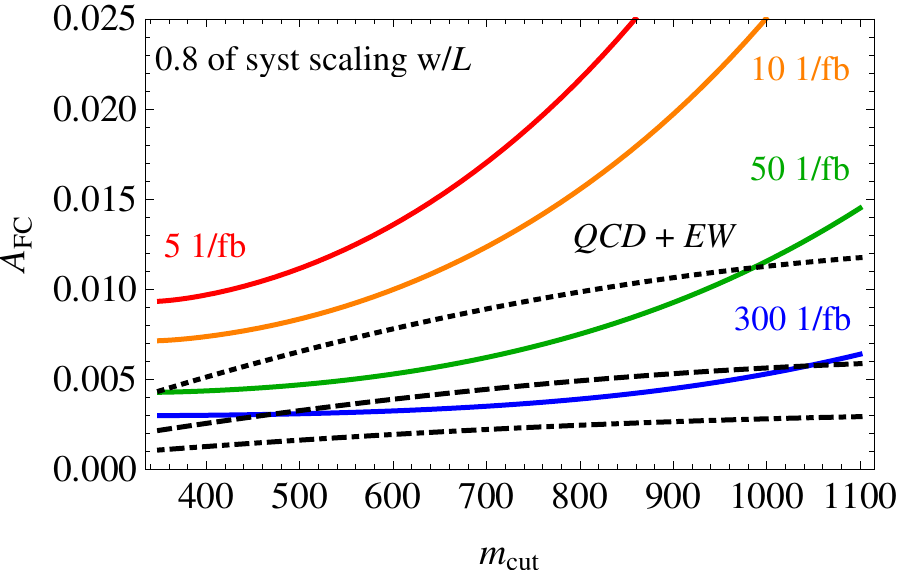}\\
\caption{ SM {\em inclusive} lepton asymmetry compared to projected
  uncertainty on the measurement at LHC 14, as a function of minimum
  $m_{t\bar t}$, with and without systematic uncertainties.  The
  dotted line shows the SM (QCD and EW combined) predictions of
  Ref.~\cite{Bernreuther:2012sx}, the dashed line shows 0.5 $\times$
  the asymmetry, and the dash-dotted line shows 0.25 the asymmetry to
  give a rough estimate of $95\%$ CL sensitivity recalling the washout
  due to finite lepton acceptance cuts. Left: statistical error only.
  Right: statistical error combined in quadrature with systematic
  error given by $\Delta A = 0.007$ at 5 fb$^{-1}$, with 0.8 of the
  systematic uncertainty scaling as $1/\sqrt{\mathcal{L}}$. }
\end{figure}
%%%%

To estimate the reach in the leptonic asymmetry we use an approximate
(flat) efficiency for top pair identification and reconstruction of
$\epsilon = 0.25$ per dileptonic top pair event, again accounting for
geometric and $p_T$ acceptance and an additional finite efficiency for
reconstructing the top pair from the measured missing momentum.
Results are shown in Fig.~\ref{fig:topkin-SMAsymmProjDilep}.
Observation of the SM effect in this channel does not look promising;
the reduction in the systematic error bars is more than compensated by
the intrinsically smaller signal.

%%%%%%%%%%%%%%%%%%%%%%%%%%%%%%%%%%%
\section{New observables for the top charge asymmetry at the LHC}
\label{sec:topkin-newObs}
%%%%%%%%%%%%%%%%%%%%%%%%%%%%%%%%%%%

In contrast to $A_{FC}$ in $t\bar{t}$ production, the asymmetry in
$t\bar{t}+j$ final states has the advantage that it is non-vanishing at LO and 
NLO calculations are available \cite{Dittmaier:2008uj,Bevilacqua:2010ve,Melnikov:2010iu,Kardos:2011qa,Alioli:2011as,Melnikov:2011qx}.
It was found that higher order corrections in the production and in the decay process 
reduce the LO asymmetry $A_{FB}$ at the Tevatron.
Refs.~\cite{Melnikov:2010iu,Melnikov:2011qx} provide arguments for these effects which should also hold for $A_{FC}$ at the LHC.
Ref.~\cite{Berge:2013xsa} introduces two new observables to access the
production-level asymmetries in $t\bar{t}+j$ events.  First, the {\it
  incline asymmetry} measures the asymmetry in the relative angle
between the decay plane (i.e., that containing the $t$, $\bar t$ and
jet), and the production plane (i.e., that defined by the jet and the
incoming partons).  Second, the {\it energy asymmetry} between the $t$
and $\bar t$ energies is particularly useful for probing asymmetries in $qg$-initiated events.  A Snowmass 2013 study for these variables was carried out
in \cite{Berge:2013csa}, and the results are briefly summarized here.

\begin{figure}
  \centering
  \label{fig:topkin-ttjasym}
\includegraphics[width=0.45\hsize,angle=270]{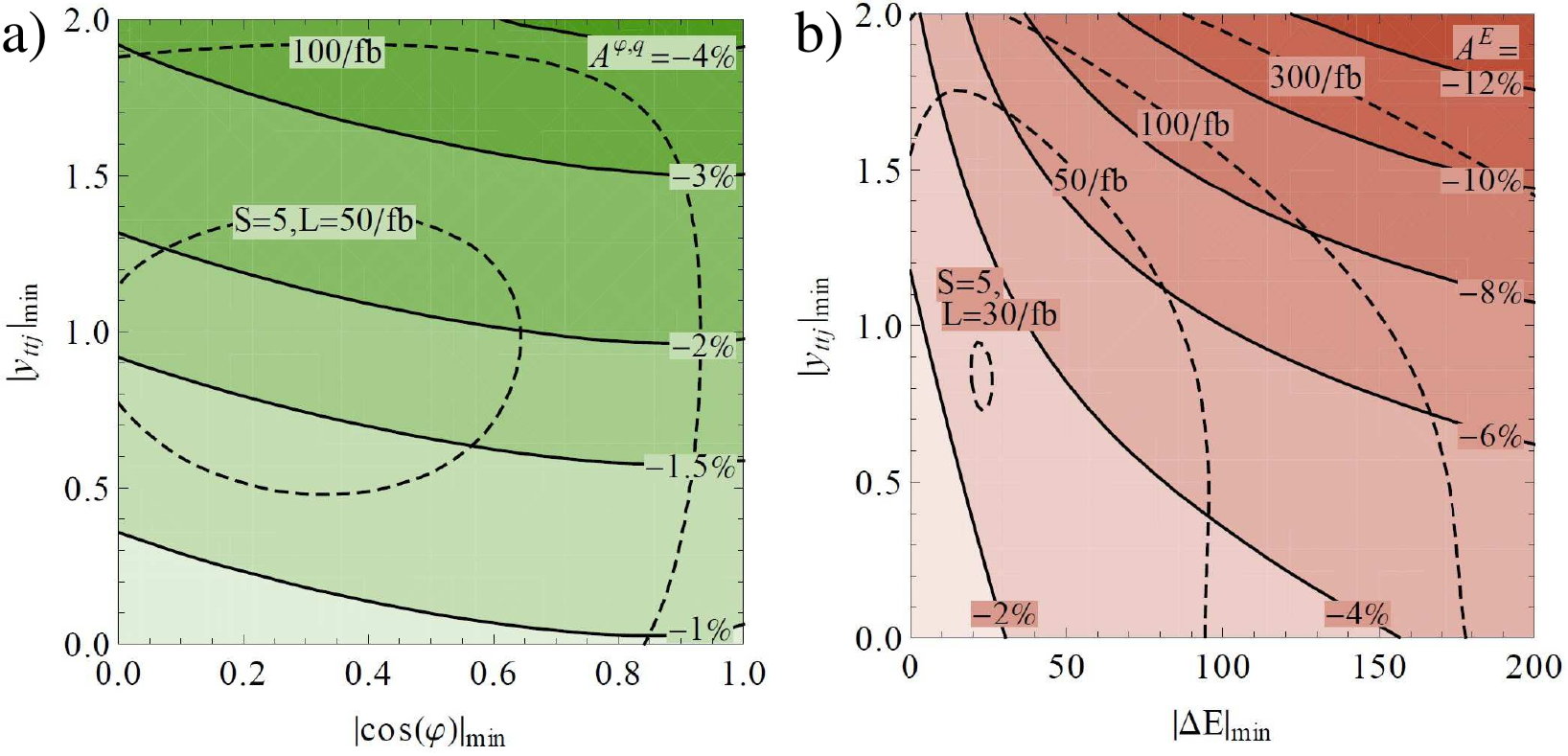}
\caption{Observability contours of asymmetry variables in $t\bar t+$jet final states at LHC14, as a function of cuts on $t\bar t +$jet events.  Left: the incline asymmetry as a function of the
  minimum required incline angle  and the minimal required
  system rapidity $|y_{t\bar{t}j}|$. Right: the energy asymmetry as a
  function of the minimum required energy difference and the minimal required system rapidity. Further details are available in Ref.~\cite{Berge:2013xsa,Berge:2013csa}.}
\end{figure}

In $t\bar{t}+j$ events, the incline asymmetry (as opposed to the
forward-central asymmetry) benefits from the additional information
provided by the $t\bar{t}+j$ event
kinematics. Ref.~\cite{Berge:2013xsa} finds that it is possible to
measure an incline asymmetry of the order of 4\% with a significance
of three standard deviations with $100~\mathrm{fb}^{-1}$ at 14~TeV
(see Fig.\ \ref{fig:topkin-ttjasym}).

%%%%
%%%%
Extrapolating this to $3000~\mathrm{fb}^{-1}$ of integrated luminosity
yields a significance of about 16 s.d.\ for both observables.Calculations carried out
for the new observables are currently performed at leading order, and
NLO corrections to the new observables could be sizeable, thus affecting
the projected sensitivities.

Because of the decreasing fraction of $q\bar{q}$-initiated events with
increasing center-of-mass energies, a measurement at 100~TeV would
require extremely strong cuts to suppress the $gg$ initial state
contributions. For reasonably large asymmetries of around 8\% wide
$|y|$, cuts of up to 5 are needed for the rapidity of the $t\bar{t}+j$ system
\cite{Berge:2013csa}.

%%%%%%%%%%%%%%
\subsection{Asymmetry at LHCb}
\label{sec:topkin-lhcb}
%%%%%%%%%%%%%

Measurement of a top-antitop rate asymmetry at the LHCb experiment
could provide further information on the top quark asymmetry
\cite{Kagan:2011yx}. LHCb has unique capabilities to measure the
charge asymmetry in top quark production. An official LHCb study was
performed for Snowmass 2013 \cite{Coco:2013} and the results are
briefly summarized here.

In top quark production at LHCb, as at ATLAS and CMS, the best reach
is provided by the lepton+jets decay channel, although the dileptonic
decay channel can provide supplementary
reach. Table~\ref{tab:lhcb_yields} summarizes cross sections as
predicted by POWHEG within the LHCb acceptance. The uncertainties are
estimated from scale variations, comparing the central PDF sets from
different groups (CTEQ, MSTW, NNPDF) and shower as well as tagging
uncertainties.  Despite lower backgrounds in the dilepton final state,
the production yield at LHCb makes measurements in the dilepton decay
channel not statistically significant, although the high luminosity
phase of the LHC may offer enough integrated luminosity to change this
conclusion.

%%%%%%%%%%%%
\begin{table}[h!]
\begin {center}
\begin{tabular}{|l|cc|}
\hline \hline
Channel & $\sigma$ (fb) at 8 TeV & $\sigma$ (fb) at 14 TeV \\ \hline
$lb$   & $675 \pm 110$ & $5353 \pm 780$  \\ 
$lbj$  & $254 \pm 41$  & $2758 \pm 410$  \\ 
$lbb$  & $111 \pm 22$  & $1296 \pm 233$  \\ 
$lbbj$ & $68 \pm 13$   & $860 \pm 156$   \\ 
$ll$   & $79 \pm 15$   & $635 \pm 109$   \\ 
$llb$  & $19 \pm 4$    & $417 \pm 79$    \\ \hline
\end{tabular}
\caption{\label{tab:lhcb_yields} Summary of $t\bar{t}$ inclusive cross section channels 
  within LHCb acceptance for LHC8 and LHC14. The predicted cross sections are computed with 
  a 60 GeV cut for the transverse momentum of the leading $b$ jet.}
\end {center}
\end{table}
%%%%%%%%%%%%%%%%%

The large pseudorapidity coverage at LHCb allows differential charge
asymmetry measurements to be made in the forward region.  Here on one
hand the intrinsic size of the asymmetry is enhanced, and on the other
hand, the dominant asymmetric background from $W+$jet production is
reduced.  The study \cite{Coco:2013} concludes that with enough 
(expected integrated luminosity after LS1 is 50 fb$^{-1}$) integrated luminosity at 14 TeV 
a measurement of the charge asymmetry in top quark production at the
LHCb can be done. The high luminosity phase of the 14 TeV LHC makes it
possible to also measure the charge asymmetry in the dilepton decay
channel.  No degradation due to pile-up effects is anticipated,
allowing unchanged trigger settings with respect to current 7 and 8
TeV running conditions.  Measurements from LHCb are complementary to
the measurements in the central region performed by ATLAS and CMS and
thus contribute to a deeper understanding of top quark production. A
combination of LHCb results with those from ATLAS and CMS would shed
further light into the top charge asymmetry puzzle.

%%%%%%%%%%%%%%%%%%%%%%%%%%%%%%%%%%%
\section{Conclusions}
\label{sec:conclusions}
%%%%%%%%%%%%%%%%%%%%%%%%%%%%%%%%%%%

We have presented an overview of the theoretical understanding of
top quark pair production, presenting a unified study of uncertainties in top kinematic distributions and surveying prospects for and uncertainties in top spin correlations.  LHC 14 has good prospects for resolving outstanding questions in
 boosted top production cross-sections and production asymmetries.   Good control of systematic
 uncertainties is necessary to decisively measure a SM-like top pair production asymmetry at LHC 14,
but other, less inclusive observables, such as the incline asymmetry of Ref.~\cite{Berge:2013xsa}
or the forward rate asymmetry at LHCb, are less limited by systematic uncertainties.

\bigskip

\noindent {\bf Acknowledgements.}  We acknowledge useful conversations
with Thorsten Chwalek and Frederic Deliot regarding sources of
systematic errors. MS is supported by US DOE under contract DE-AC02-06CH11357. JS is supported by NSF grant PHY-1067976 and
the LHC Theory Initiative under grant NSF-PHY-0969510.

%%%%%%%%%%%%%%%%%%%%%%%%%%%%%%%%%%%%%%%%%%%


\begin{thebibliography}{99}

%%
%%  bibliographic items can be constructed using the LaTeX format in SPIRES:
%%    see    http://www.slac.stanford.edu/spires/hep/latex.html
%%  SPIRES will also supply the CITATION line information; please include it.
%%


\bibitem{Coco:2013}
%  title = internal LHCb study
%  internal LHCb study, contact person V.~Coco.
   R.~Gauld,
%   title        = "Measuring top quark production asymmetries at LHCb",
   LHCb-PUB-2013-009; CERN-LHCb-PUB-2013-009 (2013)

\bibitem{Berge:2013csa} 
  S.~Berge and S.~Westhoff,
  %``Charge Asymmetry in Top Pair plus Jet Production -- A Snowmass White Paper,''
  arXiv:1307.6225 [hep-ph].
  %%CITATION = ARXIV:1307.6225;%%


%\cite{Campbell:2010ff}
\bibitem{Campbell:2010ff} 
  J.~M.~Campbell and R.~K.~Ellis,
  %``MCFM for the Tevatron and the LHC,''
  Nucl.\ Phys.\ Proc.\ Suppl.\  {\bf 205-206}, 10 (2010)
  [arXiv:1007.3492 [hep-ph]].
  %%CITATION = ARXIV:1007.3492;%%

%\cite{Bernreuther:2001rq}
\bibitem{Bernreuther:2001rq}
  W.~Bernreuther, A.~Brandenburg, Z.~G.~Si and P.~Uwer,
  %``Top quark spin correlations at hadron colliders: Predictions at next-to-leading order QCD,''
  Phys.\ Rev.\ Lett.\  {\bf 87}, 242002 (2001)
  [hep-ph/0107086].
  %%CITATION = HEP-PH/0107086;%%

%\cite{Bernreuther:2004jv}
\bibitem{Bernreuther:2004jv}
  W.~Bernreuther, A.~Brandenburg, Z.~G.~Si and P.~Uwer,
  %``Top quark pair production and decay at hadron colliders,''
  Nucl.\ Phys.\ B {\bf 690}, 81 (2004)
  [hep-ph/0403035].
  %%CITATION = HEP-PH/0403035;%%

%\cite{Frixione:2006gn}
\bibitem{Frixione:2006gn}
  S.~Frixione and B.~R.~Webber,
  %``The MC@NLO 3.3 Event Generator,''
  hep-ph/0612272.
  %%CITATION = HEP-PH/0612272;%%

%\cite{Frixione:2007nw}
\bibitem{Frixione:2007nw}
  S.~Frixione, P.~Nason and G.~Ridolfi,
  %``A Positive-weight next-to-leading-order Monte Carlo for heavy flavour hadroproduction,''
  JHEP {\bf 0709}, 126 (2007)
  [arXiv:0707.3088 [hep-ph]].
  %%CITATION = ARXIV:0707.3088;%%

%\cite{Frixione:2007nu}
\bibitem{Frixione:2007nu}
  S.~Frixione, P.~Nason and G.~Ridolfi,
  %``The POWHEG-hvq manual version 1.0,''
  arXiv:0707.3081 [hep-ph].
  %%CITATION = ARXIV:0707.3081;%%

%\cite{Melnikov:2009dn}
\bibitem{Melnikov:2009dn}
  K.~Melnikov and M.~Schulze,
  %``NLO QCD corrections to top quark pair production and decay at hadron colliders,''
  JHEP {\bf 0908}, 049 (2009)
  [arXiv:0907.3090 [hep-ph]].
  %%CITATION = ARXIV:0907.3090;%%

%\cite{Bernreuther:2010ny}
\bibitem{Bernreuther:2010ny}
  W.~Bernreuther and Z.~-G.~Si,
  %``Distributions and correlations for top quark pair production and decay at the Tevatron and LHC.,''
  Nucl.\ Phys.\ B {\bf 837}, 90 (2010)
  [arXiv:1003.3926 [hep-ph]].
  %%CITATION = ARXIV:1003.3926;%%


%\cite{Denner:2010jp}
\bibitem{Denner:2010jp}
  A.~Denner, S.~Dittmaier, S.~Kallweit and S.~Pozzorini,
  %``NLO QCD corrections to WWbb production at hadron colliders,''
  Phys.\ Rev.\ Lett.\  {\bf 106}, 052001 (2011)
  [arXiv:1012.3975 [hep-ph]].
  %%CITATION = ARXIV:1012.3975;%%

%\cite{Bevilacqua:2010qb}
\bibitem{Bevilacqua:2010qb}
  G.~Bevilacqua, M.~Czakon, A.~van Hameren, C.~G.~Papadopoulos and M.~Worek,
  %``Complete off-shell effects in top quark pair hadroproduction with leptonic decay at next-to-leading order,''
  JHEP {\bf 1102}, 083 (2011)
  [arXiv:1012.4230 [hep-ph]].
  %%CITATION = ARXIV:1012.4230;%%

%\cite{Campbell:2012uf}
\bibitem{Campbell:2012uf}
  J.~M.~Campbell and R.~K.~Ellis,
  %``Top-quark processes at NLO in production and decay,''
  arXiv:1204.1513 [hep-ph].
  %%CITATION = ARXIV:1204.1513;%%

%\cite{Denner:2012yc}
\bibitem{Denner:2012yc}
  A.~Denner, S.~Dittmaier, S.~Kallweit and S.~Pozzorini,
  %``NLO QCD corrections to off-shell top-antitop production with leptonic decays at hadron colliders,''
  JHEP {\bf 1210}, 110 (2012)
  [arXiv:1207.5018 [hep-ph]].
  %%CITATION = ARXIV:1207.5018;%%
  %17 citations counted in INSPIRE as of 18 Sep 2013



%\cite{Martin:2009iq}
\bibitem{Martin:2009iq} 
  A.~D.~Martin, W.~J.~Stirling, R.~S.~Thorne and G.~Watt,
  %``Parton distributions for the LHC,''
  Eur.\ Phys.\ J.\ C {\bf 63}, 189 (2009)
  [arXiv:0901.0002 [hep-ph]].
  %%CITATION = ARXIV:0901.0002;%%
  %1544 citations counted in INSPIRE as of 14 Jun 2013

%\cite{Ball:2012cx}
\bibitem{Ball:2012cx} 
  R.~D.~Ball, V.~Bertone, S.~Carrazza, C.~S.~Deans, L.~Del Debbio, S.~Forte, A.~Guffanti and N.~P.~Hartland {\it et al.},
  %``Parton distributions with LHC data,''
  Nucl.\ Phys.\ B {\bf 867}, 244 (2013)
  [arXiv:1207.1303 [hep-ph]].
  %%CITATION = ARXIV:1207.1303;%%
  %37 citations counted in INSPIRE as of 14 Jun 2013




%\cite{Auerbach:2013by}
\bibitem{Auerbach:2013by} 
  B.~Auerbach, S.~V.~Chekanov and N.~Kidonakis,
  %``Studies of highly-boosted top quarks near the TeV scale using jet masses at the LHC,''
  arXiv:1301.5810 [hep-ph].
  %%CITATION = ARXIV:1301.5810;%%



%\cite{Ferroglia:2013zwa}
\bibitem{Ferroglia:2013zwa} 
  A.~Ferroglia, B.~D.~Pecjak and L.~L.~Yang,
  %``Top-quark pair production at high invariant mass: an NNLO soft plus virtual approximation,''
  arXiv:1306.1537 [hep-ph].
  %%CITATION = ARXIV:1306.1537;%%


%\cite{Kuhn:2013zoa}
\bibitem{Kuhn:2013zoa} 
  J.~H.~K\"uhn, A.~Scharf and P.~Uwer,
  %``Weak Interactions in Top-Quark Pair Production at Hadron Colliders: An Update,''
  arXiv:1305.5773 [hep-ph].
  %%CITATION = ARXIV:1305.5773;%%
  %2 citations counted in INSPIRE as of 13 Jun 2013

%\cite{Baur:2006sn}
\bibitem{Baur:2006sn} 
  U.~Baur,
  %``Weak Boson Emission in Hadron Collider Processes,''
  Phys.\ Rev.\ D {\bf 75}, 013005 (2007)
  [hep-ph/0611241].
  %%CITATION = HEP-PH/0611241;%%
  %41 citations counted in INSPIRE as of 21 Jun 2013


%\cite{Abazov:2011gi}
\bibitem{Abazov:2011gi} 
  V.~M.~Abazov {\it et al.}  [D0 Collaboration],
  %``Evidence for spin correlation in $t\bar{t}$ production,''
  Phys.\ Rev.\ Lett.\  {\bf 108}, 032004 (2012)
  [arXiv:1110.4194 [hep-ex]].
  %%CITATION = ARXIV:1110.4194;%%


%\cite{ATLAS:2012ao}
\bibitem{ATLAS:2012ao} 
  G.~Aad {\it et al.}  [ATLAS Collaboration],
  %``Observation of spin correlation in $t \bar{t}$ events from pp collisions at sqrt(s) = 7 TeV using the ATLAS detector,''
  Phys.\ Rev.\ Lett.\  {\bf 108}, 212001 (2012)
  [arXiv:1203.4081 [hep-ex]].
  %%CITATION = ARXIV:1203.4081;%%

\bibitem{CMS:ttbspincorrel} 
[CMS Collaboration], report CMS-PAS TOP-12-004.


% %\cite{Melnikov:2009dn}
% \bibitem{Melnikov:2009dn} 
%   K.~Melnikov and M.~Schulze,
%   %``NLO QCD corrections to top quark pair production and decay at hadron colliders,''
%   JHEP {\bf 0908}, 049 (2009)
%   [arXiv:0907.3090 [hep-ph]].
%   %%CITATION = ARXIV:0907.3090;%%
%   %72 citations counted in INSPIRE as of 25 Jun 2013
% 
% %\cite{Bernreuther:2010ny}
% \bibitem{Bernreuther:2010ny} 
%   W.~Bernreuther and Z.~-G.~Si,
%   %``Distributions and correlations for top quark pair production and decay at the Tevatron and LHC.,''
%   Nucl.\ Phys.\ B {\bf 837}, 90 (2010)
%   [arXiv:1003.3926 [hep-ph]].
%   %%CITATION = ARXIV:1003.3926;%%
%   %86 citations counted in INSPIRE as of 25 Jun 2013


%\cite{Han:2012fw}
\bibitem{Han:2012fw} 
  Z.~Han, A.~Katz, D.~Krohn and M.~Reece,
  %``(Light) Stop Signs,''
  JHEP {\bf 1208}, 083 (2012)
  [arXiv:1205.5808 [hep-ph]].
  %%CITATION = ARXIV:1205.5808;%%
  %32 citations counted in INSPIRE as of 14 Jun 2013




%\cite{Boughezal:2012zb}
\bibitem{Boughezal:2012zb}
  R.~Boughezal and M.~Schulze,
  %``Precise predictions for top quark plus missing energy signatures at the LHC,''
  arXiv:1212.0898 [hep-ph].
  %%CITATION = ARXIV:1212.0898;%%

%\cite{Boughezal:2013xx}
\bibitem{Boughezal:2013xx} 
  R.~Boughezal and M.~Schulze, in preparation.


%\cite{Baumgart:2011wk}
\bibitem{Baumgart:2011wk} 
  M.~Baumgart and B.~Tweedie,
  %``Discriminating Top-Antitop Resonances using Azimuthal Decay Correlations,''
  JHEP {\bf 1109}, 049 (2011)
  [arXiv:1104.2043 [hep-ph]].
  %%CITATION = ARXIV:1104.2043;%%
  %8 citations counted in INSPIRE as of 14 Jun 2013

%\cite{Caola:2012rs}
\bibitem{Caola:2012rs} 
  F.~Caola, K.~Melnikov and M.~Schulze,
  %``A complete next-to-leading order QCD description of resonant $Z'$ production and decay into $t\bar t$ final states,''
  Phys.\ Rev.\ D {\bf 87}, 034015 (2013)
  [arXiv:1211.6387 [hep-ph]].
  %%CITATION = ARXIV:1211.6387;%%
  %2 citations counted in INSPIRE as of 14 Jun 2013


%\cite{Baumgart:2012ay}
\bibitem{Baumgart:2012ay} 
  M.~Baumgart and B.~Tweedie,
  %``A New Twist on Top Quark Spin Correlations,''
  JHEP {\bf 1303}, 117 (2013)
  [arXiv:1212.4888 [hep-ph]].
  %%CITATION = ARXIV:1212.4888;%%
  %4 citations counted in INSPIRE as of 14 Jun 2013

%\cite{Bernreuther:2013aga}
\bibitem{Bernreuther:2013aga} 
  W.~Bernreuther and Z.~-G.~Si,
  %``Top quark spin correlations and polarization at the LHC: standard model predictions and effects of anomalous top chromo moments,''
  arXiv:1305.2066 [hep-ph].
  %%CITATION = ARXIV:1305.2066;%%


%\cite{Chatrchyan:2012cxa}
\bibitem{Chatrchyan:2012cxa} 
  S.~Chatrchyan {\it et al.}  [CMS Collaboration],
  %``Inclusive and differential measurements of the $t \bar{t}$ charge asymmetry in proton-proton collisions at 7 TeV,''
  Phys.\ Lett.\ B {\bf 717}, 129 (2012)
  [arXiv:1207.0065 [hep-ex]].
  %%CITATION = ARXIV:1207.0065;%%



%\cite{Choudalakis:2012hz}
\bibitem{Choudalakis:2012hz} 
  G.~Choudalakis,
  %``Fully Bayesian Unfolding,''
  arXiv:1201.4612 [physics.data-an].
  %%CITATION = ARXIV:1201.4612;%%
  %1 citations counted in INSPIRE as of 26 Aug 2013


\bibitem{ATLAS-CONF-2013-078}
  G.~Aad {\it et al.} [ATLAS Collaboration],
      %``Measurement of the top quark pair production charge
      %                asymmetry in proton-proton collisions at $sqrt{s}=7$ TeV
      %                using the ATLAS detector''
     ATLAS-CONF-2013-078.

%\cite{Bernreuther:2012sx}
\bibitem{Bernreuther:2012sx} 
  W.~Bernreuther and Z.~-G.~Si,
  %``Top quark and leptonic charge asymmetries for the Tevatron and LHC,''
  Phys.\ Rev.\ D {\bf 86}, 034026 (2012)
  [arXiv:1205.6580 [hep-ph]].
  %%CITATION = ARXIV:1205.6580;%%



%\cite{ATLAS:2012sla}
\bibitem{ATLAS:2012sla} 
  [ATLAS Collaboration],
  %``Measurement of the charge asymmetry in dileptonic decay of top quark pairs in pp collisions at √s = 7 TeV using the ATLAS detector,''
  ATLAS-CONF-2012-057.
  %%CITATION = ATLAS-CONF-2012-057;%%

%\cite{CMS:iya}
\bibitem{CMS:iya} 
  [CMS Collaboration],
  %``Top charge asymmetry measurement in dileptons at 7 TeV,''
  CMS-PAS-TOP-12-010.
  %%CITATION = CMS-PAS-TOP-12-010;%%


%\cite{Dittmaier:2008uj}
\bibitem{Dittmaier:2008uj}
  S.~Dittmaier, P.~Uwer and S.~Weinzierl,
  %``Hadronic top-quark pair production in association with a hard jet at next-to-leading order QCD: Phenomenological studies for the Tevatron and the LHC,''
  Eur.\ Phys.\ J.\ C {\bf 59}, 625 (2009)
  [arXiv:0810.0452 [hep-ph]].

%\cite{Bevilacqua:2010ve}
\bibitem{Bevilacqua:2010ve}
  G.~Bevilacqua, M.~Czakon, C.~G.~Papadopoulos and M.~Worek,
  %``Dominant QCD Backgrounds in Higgs Boson Analyses at the LHC: A Study of pp -> t anti-t + 2 jets at Next-To-Leading Order,''
  Phys.\ Rev.\ Lett.\  {\bf 104}, 162002 (2010)
  [arXiv:1002.4009 [hep-ph]].

%\cite{Melnikov:2010iu}
\bibitem{Melnikov:2010iu}
  K.~Melnikov and M.~Schulze,
  %``NLO QCD corrections to top quark pair production in association with one hard jet at hadron colliders,''
  Nucl.\ Phys.\ B {\bf 840}, 129 (2010)
  [arXiv:1004.3284 [hep-ph]].

%\cite{Kardos:2011qa}
\bibitem{Kardos:2011qa}
  A.~Kardos, C.~Papadopoulos and Z.~Trocsanyi,
  %``Top quark pair production in association with a jet with NLO parton showering,''
  Phys.\ Lett.\ B {\bf 705}, 76 (2011)
  [arXiv:1101.2672 [hep-ph]].


%\cite{Alioli:2011as}
\bibitem{Alioli:2011as}
  S.~Alioli, S.~-O.~Moch and P.~Uwer,
  %``Hadronic top-quark pair-production with one jet and parton showering,''
  JHEP {\bf 1201}, 137 (2012)
  [arXiv:1110.5251 [hep-ph]].

%\cite{Melnikov:2011qx}
\bibitem{Melnikov:2011qx}
  K.~Melnikov, A.~Scharf and M.~Schulze,
  %``Top quark pair production in association with a jet: QCD corrections and jet radiation in top quark decays,''
  Phys.\ Rev.\ D {\bf 85}, 054002 (2012)
  [arXiv:1111.4991 [hep-ph]].


\bibitem{Berge:2013xsa}
  S.~Berge and S.~Westhoff
%      title          = "{Top-Quark Charge Asymmetry Goes Forward: Two New Observables for Hadron Colliders}",
  [arXiv:1305.3272 [hep-ph]].

\bibitem{Kagan:2011yx}
%  title = {Probing New Top Physics at the LHCb Experiment},
  A.~L.~Kagan, J.~F.~Kamenik, G.~Perez, S.~Stone
  Phys.\ Rev.\ Lett.\ {\bf 107},  082003 (2011)
  [arXiv:1103.3747 [hep-ph]].


\end{thebibliography}
\end{document}